\documentstyle[sprocl,psfig]{article}
\begin{document}
%\psdraft   % only draw the frame for ps figures. 

\title{NEW QCD SUM RULES FOR NUCLEON AXIAL VECTOR
COUPLING CONSTANTS}

\author{FRANK X. LEE,$^{a}$
                DEREK B. LEINWEBER,${}^{b}$
                XUEMIN JIN$\, {}^a$}
\address{
\it ${}^a$TRIUMF, 4004 Wesbrook Mall,
Vancouver, British Columbia, Canada V6T 2A3 \\
\it ${}^b$Department of Physics, University of Washington,
Seattle, WA  98195}

\maketitle

We study the predicative ability of the QCD Sum Rule method for $g_A$,
using a comprehensive Monte Carlo based uncertainty
analysis~\cite{leinweber2}.  This is the first application of such
analysis to a three-point function.

In this approach, the complete QCD parameter space is explored,
allowing a quantitative study of how the uncertainties in the QCD
input parameters propagate to the phenomenological fit parameters.
The Borel window over which the phenomenological and QCD sides of the
sum rules are matched is determined by the following criteria: a) OPE
convergence --- the last term of the truncated OPE series is less than
10\% of the total OPE side, b) ground-state dominance --- all excited
state contributions are no more than 50\% of the phenomenological
side.  Those sum rules which do not meet these criteria are considered
invalid and are discarded.

Two sets of new QCD sum rules for $g_A$ are derived in the external
field method, using generalized nucleon interpolating fields.  Three
sum rules are derived from the correlator $<\eta_{\scriptscriptstyle
1/2}\bar{\eta}_{\scriptscriptstyle 1/2}>$ of spin-1/2 interpolating
fields: $\eta_{\scriptscriptstyle 1/2}=\eta_1+\beta\,\eta_2$ where
$\beta$ is a real parameter,
$\eta_1=\epsilon^{abc}\left(u^{aT}C\gamma_5 d^b\right)u^c$, and
$\eta_2=\epsilon^{abc}\left(u^{aT}Cd^b\right)\gamma_5 u^c$.  The Ioffe
current may be recovered by setting $\beta=-1$.  However, Ioffe's
choice is not optimal.  Another set of 8 sum rules is derived
from the mixed correlator $<\eta_{\scriptscriptstyle
\mu,1/2}\bar{\eta}_{\scriptscriptstyle \nu,3/2}>$ of spin-1/2 and
spin-3/2 fields: $\eta_{\scriptscriptstyle
\mu,1/2}=\gamma_\mu\gamma_5\, \eta_{\scriptscriptstyle 1/2}$ and
$\eta_{\scriptscriptstyle \nu,3/2}=\epsilon^{abc}[
(u^{aT}C\sigma_{\rho\lambda} d^b)\sigma^{\rho\lambda} \gamma_\nu u^c
-(u^{aT}C\sigma_{\rho\lambda} u^b)\sigma^{\rho\lambda} \gamma_\nu
d^c]$, the latter has both spin-1/2 and spin-3/2 components.

\begin{figure}[htb]
\centerline{\psfig{file=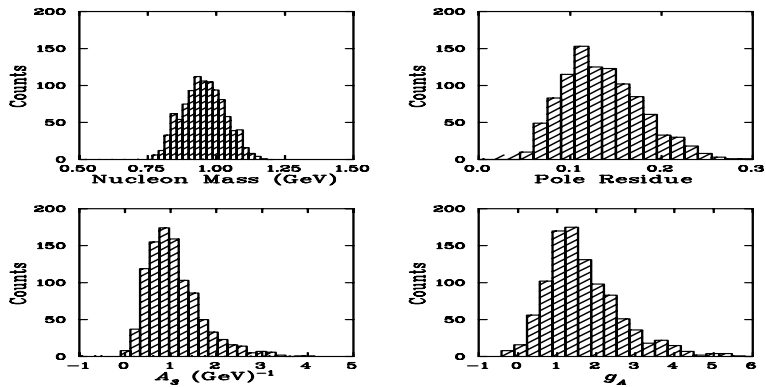,angle=90,width=10cm,height=5cm}}
\caption{Histograms of selected fit parameters drawn from 1000 sets of
QCD parameters. We find: nucleon mass $M_N=0.94 \pm 0.10$ GeV, pole
residue $\tilde{\lambda}^2_{1/2}=0.12\pm 0.6$ GeV$^6$, transition
strength $A_3=1.08\pm 0.61$ GeV$^{-1}$, and 
$g_A=1.48\pm^{1.06}_{0.65}$. \hfill\null}
\label{bin6}
\end{figure}

\begin{figure}[htb]
\centerline{\psfig{file=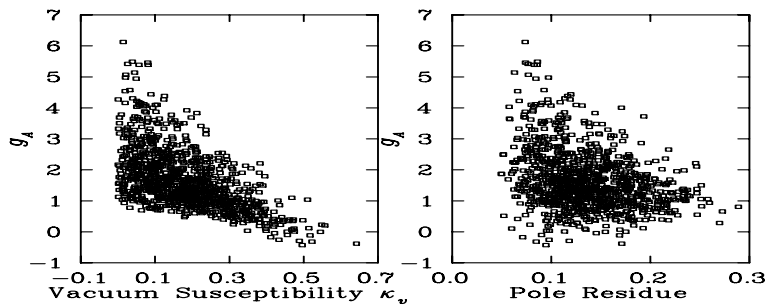,angle=90,width=10cm,height=3.9cm}}
\caption{Scatter plots of correlations with $g_A$.}
\label{corr4}
\end{figure}

Our analysis reveals that the sum rule from the spin-1/2 interpolators
which was chosen in previous works~\cite{old}, does not have a valid
Borel window.  Therefore the previous predictions for $g_A$ are
unreliable.  On the other hand, the sum rule at the structure
$iZ_\mu\sigma^{\mu\nu}p_\nu\gamma_5$ has a valid Borel window of 0.91
GeV to 1.16 GeV with the optimal mixing~\cite{leinweber2}
$\beta=-1.2$.  A combined analysis of this sum rule together with the
nucleon mass sum rules~\cite{leinweber2} yields
$g_A=1.48\pm^{1.06}_{0.65}$.  The relative error of approximately 60\%
is large compared to a 10\% error for the nucleon mass obtained from
the same input parameters.  Fig.~\ref{bin6} shows distributions for
selected fit parameters drawn from 1000 sets of QCD input parameters.
Fig.~\ref{corr4} shows two of the parameters that have significant
correlations with $g_A$.  The origin of the large error in $g_A$ is
mainly due to the poorly determined pole residue, as $g_A$ is
extracted from the form $\lambda^2\,g_A\,e^{-M_N^2/M^2}$.  The vacuum
susceptibility $\kappa_v$ also contributes to this large uncertainty.
Preliminary analysis of the sum rules from the mixed correlator shows
no improvement over this conclusion.  Analysis of other axial
couplings is under way.

This work is supported in part by the Natural Sciences and
Engineering Research Council of Canada and U.S. DOE under 
grant DE-FG06-88ER40427.

\end{document}